# AN EXAMPLE FOR THE USE OF BITWISE OPERATIONS IN PROGRAMMING

**Krasimir Yankov Yordzhev**

This piece of work presents a meaningful example for the advantages of using bitwise operations for creating effective algorithms in programming. A task connected with mathematical modeling in weaving industry is examined and computed.

2000 MSC: 68N15, 68R05, 05A18

Key words: C/C++ programming language, bitwise operation, binary matrix, equivalence relation, factor set, weaver structure, binary system, lexicographic order

**1. Introduction.** The use of bitwise operations is a powerful method used in C/C++ programming languages. Unfortunately in widespread books on this topic there is incomplete or no description for the work of the bitwise operations [2,4,5,9,11]. The aim of this article is to correct this lapse to a certain extent and present a meaningful example of a programming task, where the use of bitwise operations is appropriate in order to facilitate the work and to increase the effectiveness of the respective algorithm.

On the other hand the algorithm specified here could have a good practical application for computing a known combinatorial task connected with the classification of the various textile structures.

**2. Task formulation.** Let us denote by $Bn$ the set of all $n \times n$ binary matrices, i.e. matrices composed by $n$ rows and $n$ columns, all elements of which are either 0 or 1. It's a well-known fact that the number of all matrices of $Bn$ is equal to $2^{n^2}$. Let $A, B \in Bn$. We will say, that $A$ and $B$ are equivalent and we will write $A \sim B$, if $B$ is obtained from $A$ as a result of sequential cyclic move of the last row or column at a first place. It's easy to see that the so described relation is an equivalence relation. So this way we come to a formulation of the following programming task:

**Task 1.** *Write a program that with assigned positive integer $n$ returns one representative of each equivalence class in $Bn$ concerning the above mentioned equivalence relation.*

As a result from the solution of task 1 we will also compute a combinatorial task to find the number of all equivalence classes in $Bn$ regarding the equivalence relation $\sim$, i.e. for finding the cardinal number of the factor set $Bn_{/\sim}$.

This task is applicable in wavering industry. With the help of the elements of $Bn$ the various threads interweaving of a certain weaver structure could be coded, and with this coding by using two equivalent matrices the weaving of one and the same fabric is coded, because of cyclic recurrence of the repetition of interweaving [6,8].

From a practical point of view just matrices with at least one 0 and at least one 1 in each row and each column have meaning. Let's mark with $Qn$ the set of all matrices of that kind, $Qn \subset Bn$. The next task which we are going to compute is a bit more difficult version of task 1.

**Task 2.** *Write a program that with assigned positive integer $n$ returns one representative of each equivalence class in $Q_n$ concerning the above mentioned equivalence relation.*

**3. Bitwise operations in C/C++.** Bitwise operations can be applied for integer data type only, i.e. they cannot be used for float and double types. For the definition of the bitwise operations in C/C++ and some of their elementary applications could be seen, for example, in [1,3,7,10].

We assume as usually that bits numbering in variables starts from right to left, and that the number of the very right one is 0.

Let $x$, $y$ and $z$ are integer variables of one type, for which $w$ bits are needed. Let $x$ and $y$ are initialized and let the $z = x\alpha y$ assignment is made, where $\alpha$ is one of the operators & (**bitwise AND**), | (**bitwise inclusive OR**) or ^ (**bitwise exclusive OR**). For each $i = 0,1,\ldots,w-1$ the new contents of the $i$ bit in $z$ will be as it is presented in the following table:

| The $i$ bit of $x$ | The $i$ bit of $y$ | The $i$ bit of $x$ & $y$ | The $i$ bit of $x \mid y$ | The $i$ bit of $x$^$y$ |
|---|---|---|---|---|
| 0 | 0 | 0 | 0 | 0 |
| 0 | 1 | 0 | 1 | 1 |
| 1 | 0 | 0 | 1 | 1 |
| 1 | 1 | 1 | 1 | 0 |

In case that $z =\sim x$, if the $i$ bit of $x$ is 0, then the $i$ bit of $z$ becomes 1, and if the $i$ bit of $x$ is 1, then the $i$ bit of $z$ becomes 0, $i = 0,1,\ldots,w-1$.

In case that $k$ is a nonnegative integer, then the statement $z = x << k;$ (**bitwise shift left**) will write in the $(i+k)$ bit of $z$ the value of the $k$ bit of $x$, where $i = 0,1,\ldots,w-k-1$, and the very right $k$ bits of $z$ will be filled by zeroes. This operation is equivalent to a multiplication of $x$ by $2^k$. The statement $z = x >> k$ works the similar way (**bitwise shift right**). But we must be careful here, as in various programming environments (see for example in [7]) this operation has different interpretations – somewhere $k$ bits of $z$ from the very left place are compulsory filled by 0 (logical displacement), and elsewhere the very left $k$ bits of $z$ are filled with the value from the very left (sign) bit; i.e. if the number is negative, then the filling will be with 1 (arithmetic displacement). Therefore it's recommended to use unsigned type of variables (if the opposite is not necessary) while working with bitwise operations.

Directly form the definition of the operation bitwise shift left follows the effectiveness of the following function computing $2^k$, where $k$ is a nonnegative integer:

```
unsigned int Power2(unsigned int k) {
     return 1<<k;
}
```

To compute the value of the $i$ bit of an integer variable $x$ we can use the function:

```
int BitValue(int x, unsigned int i) {
      if ( (x & (1<<i) ) == 0 ) return 0;
      else return 1;
}
```

Bitwise operations are left associative.

The priority of operations in descending order is as follows: *bitwise complement* ~; *arithmetic operations* * (multiply), / (divide), % (remainder or modulus); *arithmetic operations* + (binary plus or add) - (binary minus or subtract); the *bitwise operations* << and >>; *relational operations* <, >, <=, >=, ==, !=; *bitwise operations* &,^ and |; *logical operations* && and ||.

**4. Algorithm realization.** Each $n \times n$ binary matrix $A$ can be coded with the help of vector (array) of $n$ nonnegative integers $v = (v_0, v_1, \ldots, v_{n-1})$, where $0 \leq v_i \leq 2^n - 1$ for each $i$: $0 \leq i \leq n-1$. One-to-one correspondence is realized through binary presentation of natural numbers, i.e. the $i$ row of the matrix $A$ is $v_i$ in binary system. The row $i$ of $A$ will be completely nil if and only if $v_i = 0$; and all elements of the $i$ row of $A$ will be equal to 1 if and only if $v_i = 2^n - 1$. In other words, it's a necessary and sufficient condition for each $i = 0, 1, \ldots, n-1$ to be realized $1 \leq v_i \leq 2^n - 2$, in order to obtain at least one 0 and at least one 1 in each row. In order to obtain at least one 0 in each column of the matrix $A$, it is necessary and sufficient that the bitwise AND of all numbers, representing the rows of $A$ to be equal to 0. In order to obtain at least one 1 in each column of the matrix $A$ it is necessary and sufficient that the bitwise inclusive OR of all numbers, representing the rows of $A$ to be equal to $2^n - 1$, i.e. to be equal to a number which is written in binary system with exact $n$ number of 1 and not even one 0.

Thus we obtain the following function, which checks whether the array of $n$ integers $v = (v_0, v_1, \ldots, v_{n-1})$ represents a matrix of $Qn$, or not

```
int IsQn(unsigned int v[], unsigned int n) {
  // Returns 1, if with v a matrix in Qn is coded
  // Returns 0, otherwise
   for (int i=0; i <= n-1; i++)
      if  (v[i]<1 || v[i] > (1<<n)-2) return 0;
   int x,y;
   x = (1<<n) -1;
   y=0;
   for (int j=0; j <= n-1; j++) {
      x = x & v[j];
      y = y | v[j];
   }
   if (x != 0) return 0;
   if (y != (1<<n)-1) return 0;
return 1;
}
```

Let $x$ be an integer, for which we are certain that it belongs to the interval $0 \leq x \leq 2^n - 1$, i.e. there's no need of more than $n$ digits 0 or 1 for its binary code. Then to present $x$ in binary system (see the function BitValue described in the previous section), written with the aid of exactly $n$ digits 0 or 1 and eventually with a certain number of insignificant zeroes at the beginning, we can use the following function:

```
void BinPrn(int x, unsigned int k) {
   int z;
   for (int i = k-1; i >= 0; i--) {
      z = x & (1<<i);
      if (z == 0) cout<<'0';
           else cout<<'1';
      }
    cout<<'\n';
}
```

Let us examine the set
$$V = \{(v_0, v_1, \ldots, v_{n-1}) \mid 0 \leq v_i \leq 2^n - 1, i = 0, 1, \ldots, n-1\} .$$

All elements of $V$ can be sorted in ascending lexicographic order. The essence of the proposed by us algorithm is to obtain sequentially all elements of $V$ in the same increasing order from the smallest one to the biggest one and right after obtaining them to check whether this element is minimal according to the lexicography order in he class of equivalence. At last we will separate just the minimal in their class of equivalency elements and they will be the only representatives of each equivalent class in the sets $Bn$ and $Qn$ (which was required in Tasks 1 and 2). For this purpose we will design function IsMin, which will return 1, if the input argument is minimal in the class of equivalency to which it belongs to, and 0 otherwise. But before that we need the following auxiliary function CicleMove, which from assigned nonnegative integers $x$ and $n$ return a number, which is obtained from $x$ by moving all bits with one to the right, beginning with the moving of the very right bit to the place of the bit $n-1$. In this case we will be helped by the bitwise operations.

```
unsigned int CicleMove(unsigned int x,unsigned int n) {
   unsigned int b0 = x & 1;    // Record the value
                    //of the very right bit of x
   x=x & ((1<<n)-1);   // Replaces all bits to the
      //left from the on with number (n-1) with 0
   return (x >> 1)|(b0 << n-1);
}
```

The following auxiliary function will also be useful for the computing of the main task:

```
Int IsLess(unsigned int u[],unsigned int v[],int n)
{
// Return 1, if according to lexicographic order
    //u[0] u[1] … u[n-1] < v[0] v[1] … v[n-1]
// Return 0, otherwise
```

```
    int i = 0;
  while ((u[i] == v[i]) && (i<n-1)) i++;
    if (u[i] < v[i]) return 1;
       else return 0;
}
```

The above mentioned function IsMin could look as follows:

```
int IsMin(unsigned int v[], unsigned int n) {
// Return 1, if according to lexicographic order
//v[0] v[1] … v[n-1]is minimal in its class of
                                    // equivalency
// Return 0, otherwise
   unsigned int u[32], v1[32];
   for (int i = 0; i <= n-1; i++)    v1[i] = v[i];
   for (int j = 0; j <= n-1; j++) {
      for (int i = 1; i <= n-1; i++) {
           for (int s = 0; s <= n-1; s++) {
                int s1 = (s+i) % n;
                u[s] = v1[s1];
           }
           if (IsLess(u,v,n) ) return 0;
      }
      for     (int    i=0;   i    <=    n-1;    i++)
v1[i]=CicleMove(v1[i],n);
   }
      return 1;
}
```

Taking the advantages of the above described functions we propose the following computing of tasks 1 and 2 (for n=4, for example). In order to be brief here we will not print all the elements obtained, and we will obtain their number only. For the hard-copy itself for each row of any of the obtained matrices we can take advantage of the above described procedure BinPrn and after organizing of a cycle by the number of the row to print the whole matrix as well. .

```
int main() {
  const int n=4;
  int i;
  unsigned long int NBn = 0;   // Number of elements
                                    //in Bn
  unsigned long int NQn = 0;   // Number of elements
                                    //in Qn
  unsigned long int NBnEq = 0;  // Number of the
                     //classes of equivalency in Bn
  unsigned long int NQnEq = 0;  // Number of the
                     //classes of equivalency in Qn
  unsigned int v[n];
  int r=(1<<n)-1;
  for (i = 0; i<n; i++)  v[i]=0;
  do {
    i=n-1;
    for (int k=0; k<=r; k++) {
       v[i]=k;
       NBn++;
```

```
            if (IsQn(v,n) ) NQn++;
            if (IsMin(v,n) ) {
                NBnEq++;
                if (IsQn(v,n))   NQnEq++;
           }
      }
    }
    while (v[i]==r) i--;
     if (i>=0) {
        v[i]++;
        for (int k=n-1; k>i; k--) {
             v[k]=0;
        }
     }
  } while ( i>=0 );
  cout<<"Number of elements in Bn "<<NBn<<'\n';
  cout<<"Number of elements in Qn "<<NQn<<'\n';
  cout<<"Number of the classes of equivalency in Bn
"<<NBnEq<<'\n';
  cout<<"Number of the classes of equivalency in Qn
"<<NQnEq<<'\n';
  return 0;
}
```

The results from the above described program for some values of $n$ can be summarized in the following table.

| $n$ | 1 | 2 | 3 | 4 | 5 | 6 |
|---|---|---|---|---|---|---|
| $\|Bn\|$ | 2 | 16 | 512 | 65 536 | 33 554 432 | $2^{36} > 2^{32} - 1$ |
| $\|Qn\|$ | 0 | 2 | 102 | 22 874 | 17 633 670 | $> 2^{32} - 1$ |
| $\|Bn_{/\sim}\|$ | 2 | 7 | 64 | 4 156 | 1 342 208 | 1 908 897 152 |
| $\|Qn_{/\sim}\|$ | 0 | 1 | 14 | 1 446 | 705 366 | 1 304 451 482 |

Krasimir Yankov Yordzhev
South-West University "N. Rilsky"
2700 Blagoevgrad, Bulgaria
e-mail: iordjev@swu.bg, iordjev@yahoo.com